\documentclass[12pt, a4paper]{article}
\usepackage[utf8]{inputenc}
\usepackage{hyperref}
\usepackage{geometry}
\usepackage{graphicx}
\usepackage{subcaption}
\usepackage[numbers]{natbib}
\usepackage{amssymb}                                                                                                                                                                                                                                                                                                                                                                                                                                                                                                                                                                                                                                                                                                                                                                                                                                                                                                                                                                                                                                                                                                                                                                                                                                                                                                                                                                                                                                                                                                                                                                                                                                                                                                                                                                                                                                                                                                                                                                                                                                                                                                                                                                                                                                                                                                                                                                                                                                                                                                                                                                                                                                                                                                                                                                                                                                                                                                                                                                                                                                                                                                                                                                                                                                                                                                                                                                                                                                                                                                                                                                                                                                                                                                                                                                                                                                                                                                                                                                                                                                                                                                                                                                                                                                          
\usepackage{amsmath}
\usepackage{slashed}
\usepackage{anyfontsize}
\usepackage{bm}
\usepackage{slashed}
\usepackage{multirow}
\usepackage{float}
\usepackage[font=small]{caption}
\usepackage{listings}
\newgeometry{lmargin=1.1in, rmargin=1.1in, tmargin=1in, bmargin=1in}
\title{Quantization of Generalized Abelian Gauge Field Theory under Rotor Model}
\author{B.T.T.Wong\footnote{CERN, u3500478@connect.hku.hk}}
\date{}

\begin{document}

\maketitle
\begin{abstract}
This paper is a follow-up work of the previous study of the generalized abelian gauge field theory under rotor model of order $n$ of higher order derivatives. We will study the quantization of this theory using path integral approach and find out the Feynman propagator (2-point correlation function) of this generalized theory.  We also investigate the generalized Proca action under rotor model and derive the Feynman propagator for the massive case. 
\end{abstract}

\section{Introduction}
Higher order derivative quantum field theory, as an extension of conventional second-order theory, is of particular interest because of its power of eliminating ultraviolet (UV) divergences in scattering amplitudes \citep{ho1,ho2,ho3,ho4,ho5}. However, there are problems of renormalizability of these theories \cite{ho6, ho6a}. There are studies on higher order derivative scalar field and gauge field theories, and these theories show contributions in quantum gravity and modified gravity \citep{h1,h2,h3,h4,h5, ho6b, ho6c, ho8}. Higher order derivative theory also appears in non-local theory such as string theory \citep{h9,h10,h11,h12,h13}. Quantization of higher order derivative quantum field theory using path integral approach has been studied in \citep{ho9, ho10, ho11,ho12,ho13}.

In our previous work in \citep{BW}, we have proved a theorem of generalized renormalizable abelian Maxwell action under rotor model with higher order derivatives,
\begin{equation}
S = -\frac{1}{4} \int d^D x G_{n\,\mu\nu} G^{\mu\nu}_n  = \frac{1}{4^n}\int d^D x \big( \Box^n T^{\mu} \big) \hat{R}_{\mu\nu}  \big( \Box^n T^{\nu} \big) =- \frac{1}{4^{n+1}} \int d^{D}x \,\Box^n G_{\mu \nu} \Box^n G^{\mu \nu}\,,
\end{equation}
where $\hat{R}_{\mu\nu}$ is the projection tensor $\hat{R}_{\mu\nu} = \frac{1}{2}( \Box \eta_{\mu\nu} - \partial_{\mu}\partial_{\nu}) $ and $G_{n\,\mu\nu}= \partial_{\mu}T_{n\,\nu}-\partial_{\nu}T_{n\,\mu} $ is the field strength of the $n^\mathrm{th}$ order rotor gauge field strength. The $n$-th order gauge field strength is identified as \citep{BW}
\begin{equation}
G_{n\,\mu\nu} \equiv \frac{1}{2^n} \Box^n G_{\mu\nu} \,.
\end{equation}

As the projection is regarded as second order rotation \citep{BW}, the $n$-th order rotation of gauge field $T_{n}^{\mu_n}$ is attained by successive second order rotation as \citep{BW}
\begin{equation}
T_{n}^{\mu_n} = \hat{R}_{\mu_n \mu_{n-1}}\hat{R}^{\mu_{n-1} \mu_{n-2}} \cdots \hat{R}_{\mu_3 \mu_2} \hat{R}^{\mu_2 \mu_1} \hat{R}_{\mu_1 \mu_0} T^{\mu_0} = \frac{1}{2^{n-1}}  P_{\mu_n}^{\,\,\,\mu_{n-1}} P_{\mu_{n-1}}^{\,\,\,\mu_{n-2}} \cdots P_{\mu_3}^{\,\,\,\mu_{2}} P_{\mu_2}^{\,\,\,\mu_{1}} \hat{R}_{\mu_1 \mu_0} T^{\mu_0} \,,
\end{equation}
where \begin{equation}
P_{\mu_j}^{\,\,\,\mu_{j-1}} = \Box \delta_{\mu_j}^{\,\,\,\mu_{j-1}}\,,
\end{equation}
is the propagator. In the generalized theory, the action changes by the transformation of gauge field as  $T^{\mu} \rightarrow \Box^n T^{\mu}$. The working dimension $D$ for renormalizability is $4n+4$ for unity gauge field dimension \citep{BW}. When $n=0$ this reduces to the conventional Maxwell action
\begin{equation} \label{eq:AbelianGaugeAction}
S = -\frac{1}{4} \int d^4 x G_{\mu\nu} G^{\mu\nu} = \int d^4 x T^{\mu}\hat{R}_{\mu\nu} T^{\nu} \,.
\end{equation}
The equation of motion of this theory can be obtained by minimizing the action
\begin{equation}
\frac{\delta S}{\delta (\Box^n T^\mu)} = 0 \,,
\end{equation}
then we obtain the equation of motion as \citep{BW}
\begin{equation}
\hat{R}_{\mu\nu} \Box^n T^{\nu} = 0 \,.
\end{equation}
The Noether's conserved current is \citep{BW}
\begin{equation}
     J^{\mu} = \Box^n G^{\nu\mu} \partial_{\nu}\Box^n  \theta \,,
\end{equation}
and the associated Noether's charge $Q$ is \citep{BW}
\begin{equation}
 Q = \int d^{D-1} x J^{0} = \int d^{D-1} x \Box^n G^{\nu 0} \partial_{\nu}\Box^n  \theta \,.
\end{equation}
When $n=0$, it restores back to the conventional Maxwellian case. 

Quantization of the generalized abelian gauge field theory under rotor model can be proceeded with Feynman path integral approach. The quantum amplitude can be computed as an integral of all possible field configurations over the exponential of the action \citep{Feyn1,Feyn2,Peskin}. Since in the generalized model it involves the transformation of field by $T^{\mu} \rightarrow \Box^n T^{\mu}$, therefore in the path integral we sum over all possible configurations of $\Box^n T^{\mu}$ instead of $T^{\mu}$, i.e. the integration measure changes by
\begin{equation}
\int \mathcal{D}T^{\mu}(x) \rightarrow \int \mathcal{D}\Box^n T^{\mu}(x) \,.
\end{equation}

\section{Path integral quantization of generalized abelian gauge field theory}
In this section, we will study the quantization of general abelian gauge field theory by path integral approach in detail. From now on, we take the transformed $\Box^n T^\mu$ field as field variable. Simply speaking, the physics is changed by $T^{\mu}\rightarrow \Box^n T^{\mu}$. The quantum amplitude of the $\Box^n T^\mu$ field in the renormalizable $4n+4$ dimension is
\begin{equation}
\langle \Box^n T^{\mu}_f (t_f, \pmb{\mathrm{x}}) | e^{-i\hat{H}(t_f - t_i )} | \Box^n T^{\mu}_i ( t_i , \pmb{\mathrm{x}} ) \rangle = \int \mathcal{D} \Box^n T^{\mu}(x) \, \exp\bigg(-\frac{i}{4^{n+1}}\int d^{4n+4}x \Box^n G_{\mu\nu} \Box^n G^{\mu\nu}\bigg) \,,
\end{equation}
where $| \Box^n T^{\mu}_i ( t_i , \pmb{\mathrm{x}} ) \rangle$ is the field state at initial time $t_i$ and $| \Box^n T^{\mu}_i ( t_f , \pmb{\mathrm{x}} ) \rangle$ is the field state at final time $t_f$, and $\hat{H}$ is the Hamiltonian.
The sourced generating functional is a functional of 4-(covariant) transformed vector current $\Box^n J_{\mu}(x)$,
\begin{equation}
Z[\Box^n J_{\mu}(x)] = \int \mathcal{D}\Box^n T^{\mu}(x)\,\exp\bigg(i\int d^{4n+4} x \big(\frac{-1}{4^{n+1}} \Box^n G_{\mu\nu} \Box^n G^{\mu\nu}   \big) \,+\, i \int d^{4n+4} x \Box^n J_{\mu}(x) \Box^n T^{\mu}(x) \bigg) \,. 
\end{equation} \label{eq:FullPartitionFunctionalGaugeField}
The normalized generating functional for gauge field is
\begin{equation}
\mathcal{Z}[\Box^n J_{\mu}] = \frac{Z[\Box^n J_{\mu}]}{Z[0]} = \frac{\int \mathcal{D}\Box^n T^{\mu}\,\exp\Big(i\int d^{4n+4} x \big(\frac{-1}{4^{n+1}} \Box^n G_{\mu\nu}\Box^n G^{\mu\nu} \big)  \,+\, i \int d^{4n+4} x \Box^n J_{\mu} \Box^n T^{\mu} \Big) }{\int \mathcal{D}\Box^n T^{\mu}\,\exp\big(i\int d^{4n+4} x \frac{-1}{4^{n+1}} \Box^n G_{\mu\nu} \Box^n G^{\mu\nu}  \big) }
\end{equation}

The functional derivative on the current also involves the Kronecker delta on the Lorentz indices,
\begin{equation}
\frac{\delta \Box^n J_{\mu}(x)}{\delta \Box^n J_{\nu}(y)} = \delta^{\nu}_{\mu}\delta^4 (x-y) \,\,\,\, \text{or}\,\,\,\, \frac{\delta \Box^n J^{\mu}(x)}{\delta \Box^n J^{\nu}(y)} = \delta^{\mu}_{\nu} \delta^4 (x-y) \,.
\end{equation}
The free $n$-point correlation function is given by
\begin{equation}
\langle 0 | \mathrm{T} \Box^n \hat{T}^{\nu_1}(x_1)  \Box^n\hat{T}^{\nu_2}(x_2) \cdots \Box^n \hat{T}^{\nu_n}(x_n) |0 \rangle = \frac{1}{i^n} \frac{\delta^n}{\delta\Box^n J_{\nu_1}(x_1) \delta \Box^n J_{\nu_2}(x_2) \cdots \delta \Box^nJ_{\nu_n}(x_n)} \mathcal{Z}[\Box^n J_{\mu}] \bigg\vert_{\Box^n J_{\mu} =0} \,,   
\end{equation}
where noting that each gauge field has different Lorentz indices and $\mathrm{T}$ means the time ordering operator. Then the path integral representation is
\begin{equation} \label{eq:GaugeFieldPathInteral}
\begin{aligned}
& \quad \langle 0 | \mathrm{T} \Box^n \hat{T}^{\nu_1}(x_1)   \cdots \Box^n \hat{T}^{\nu_n}(x_n) |0 \rangle \\
&= \frac{\int \mathcal{D}\Box^n T^{\mu}(x)\,\Box^n T^{\nu_1}(x_1)  \cdots \Box^n T^{\nu_n}(x_n)   \exp\big(i\int d^{4n+4} x \frac{-1}{4^{n+1}} \Box^n G_{\mu\nu}\Box^n G^{\mu\nu}  \big) }{\int \mathcal{D}\Box^n T^{\mu}(x) \,\exp\big(i\int d^{4n+4} x \frac{-1}{4^{n+1}} \Box^n G_{\mu\nu}\Box^n G^{\mu\nu} \big)} \,.
\end{aligned}
\end{equation}

Next, we want to calculate the explicit form of Feynman propagator which is the two-point correlation function of the $\Box^n T^{\mu}$ fields
\begin{equation}
\langle 0 | T \Box^n T^{\mu} (x) \Box^n T^{\nu} (y) |0 \rangle = B^{\mu\nu}(x-y) \,.
\end{equation}
First, consider using integration by parts, i.e. the second equation in equation (1), we write the sourced generating functional as 
\begin{equation} \label{eq:keyEq1}
\begin{aligned}
Z[\Box^n J_{\mu}(x)] = \int \mathcal{D}\Box^n T^{\mu}(x)\,\exp\bigg(\frac{i}{2}\iint d^{4n+4} x \,d^{4n+4} y \,\Box^n T^{\mu}(x)\delta^{4n+4}(x-y)\big(\frac{1}{4^n}\hat{K}_{\mu\nu}\big) \Box^n T^{\nu}(y)    \\
\quad\quad\quad\quad\,+\, i \int d^{4n+4} x \Box^n J_{\mu}(x) \Box^n T^{\mu}(x) \bigg) \,,
\end{aligned} 
\end{equation} 
where $\hat{R}_{\mu\nu}=\frac{1}{2}\hat{K}_{\mu\nu}$. And the sourceless partition function is 
\begin{equation} \label{eq:keyEq2}
Z[0] = \int \mathcal{D}\Box^n T^{\mu}(x)\,\exp\bigg(\frac{i}{2}\iint d^{4n+4} x \,d^{4n+4} y \,\Box^n T^{\mu}(x)\delta^{4n+4}(x-y)\big(\frac{1}{4^n}\hat{K}_{\mu\nu}\big) \Box^n T^{\nu}(y) \bigg).  
\end{equation}
Then we use several identities from Gaussian path integral. For the first identity \citep{Peskin,Nair, Weinberg},
\begin{equation}
Z[0]= \int \mathcal{D} f(x) \exp\bigg( \frac{i}{2}\iint dx dx^\prime f(x) \hat{A}(x, x^\prime) f(x^\prime)  \bigg) = \lim_{N \rightarrow \infty}\sqrt{\frac{(2\pi i)^N}{\mathrm{det}\,\hat{A}(x,x^\prime)}}\,.
\end{equation}
For the second identity  \citep{Peskin,Nair, Weinberg},
\begin{equation}
\begin{aligned}
Z[b(x)]&=\int \mathcal{D} f(x) \exp\bigg( \frac{i}{2}\iint dx dx^\prime f(x) \hat{A}(x, x^\prime) f(x^\prime)  + i \int dx b(x)f(x) \bigg) \\
&= \lim_{N \rightarrow \infty}\sqrt{\frac{(2\pi i)^N}{\mathrm{det}\,\hat{A}(x,x^\prime)}}\exp\bigg(-\frac{i}{2} \iint dx dx^\prime b(x) \hat{A}^{-1}(x,x^\prime) b(x^\prime) \bigg) \\
&= Z[0]\exp\bigg(-\frac{i}{2} \iint dx dx^\prime b(x) \hat{A}^{-1}(x,x^\prime) b(x^\prime) \bigg) \,.
\end{aligned}
\end{equation}
And the normalized generating functional is
\begin{equation}
\mathcal{Z}[b(x)] = \frac{Z[b(x)]}{Z[0]} = \exp\bigg(-\frac{i}{2} \iint dx dx^\prime b(x) \hat{A}^{-1}(x,x^\prime) b(x^\prime) \bigg) \,.
\end{equation}
We substitute $f(x) \rightarrow \Box^n T^\mu (x)$, $b(x) \rightarrow \Box^n J_{\mu}(x)$, $\hat{A}(x,y) \rightarrow \frac{1}{4^n}\delta^{4n+4}(x-y) \hat{K}_{\mu\nu} $, we have for equation (\ref{eq:keyEq2}) as
\begin{equation}
Z[0] =   \lim_{N \rightarrow \infty}\sqrt{\frac{4^n(2\pi i)^N}{\mathrm{det}\,[ \delta^{4n+4}(x-y)(\Box \eta_{\mu\nu} -\partial_{\mu}\partial_{\nu} )]}}\,.
\end{equation}
And we have for the sourced generating functional as
\begin{equation}
Z[\Box^n J^{\mu}(x)] = Z[0] \exp\bigg(-\frac{i}{2} \iint d^{4n+4}x\,d^{4n+4}y\,\Box^n J^{\mu}(x)\cdot 4^n(\hat{K}^{-1})_{\mu\nu} \Box^n J^{\nu}(y)\bigg)\,.
\end{equation}
The normalized generating functional is
\begin{equation}
\mathcal{Z}[\Box^n J^{\mu}(x)] = \frac{Z[\Box^n J^{\mu}(x)]}{Z[0] } = \exp\bigg(-\frac{i}{2} \iint d^{4n+4}x\,d^{4n+4}y\,\Box^n J^{\mu}(x)\cdot 4^n(\hat{K}^{-1})_{\mu\nu} \Box^n J^{\nu}(y)\bigg) \,.
\end{equation}
However, the projection tensor is not invertible, this is due to the consequence of gauge invariance. We need to fix the gauge. The gauge-fixed action is
\begin{equation}
S_{\mathrm{GF}}= \int d^{4n+4} x \bigg(-\frac{1}{4^{n+1}}  \Box^n G_{\mu\nu} \Box^n G^{\mu\nu} -\frac{1}{2 \cdot 4^{n}}( \partial_{\mu}\Box^n A^{\mu})^2  \bigg)\,.
\end{equation}
Using integration by parts,
\begin{equation} \label{eq:step}
\begin{aligned}
S_{\mathrm{GF}} &=\frac{1}{2\cdot 4^n}\int d^{4n+4} x \bigg( \Box^n T^{\mu}(x)\big(\Box\eta_{\mu\nu}-\partial_{\mu}\partial_{\nu} \big)\Box^n T^{\nu}(x) + \Box^n T^{\mu}(x) \partial_{\mu}\partial_{\nu} \Box^n T^{\nu}(x) \bigg) \\
&= \frac{1}{2}\int d^{4n+4} x \Box^n T^{\mu}(x)\big(\frac{1}{4^n}\Box \eta_{\mu\nu} \big) \Box^n T^{\nu}(x) \\
&=\frac{1}{2}\int d^{4n+4} x \Box^n T^{\mu}(x) \hat{M}_{\mu\nu} \Box^n T^{\nu}(x) \,,
\end{aligned}
\end{equation}
where we define
\begin{equation}
\hat{M}_{\mu\nu} = \frac{1}{4^n}\Box \eta_{\mu\nu} \,.
\end{equation}
Then the gauge-fixed generating functional is 
\begin{equation}
\begin{aligned}
Z_{\mathrm{GF}}[\Box^n J_{\mu}(x)] = \int \mathcal{D}\Box^n T^{\mu}(x)\,\exp\bigg(\frac{i}{2}\iint d^{4n+4} x \,d^{4n+4} y \,\Box^n T^{\mu}(x)\delta^{4n+4}(x-y)\big(\frac{1}{4^n}\Box \eta_{\mu\nu}\big) \Box^n T^{\nu}(y)    \\
\quad\quad\quad\quad\,+\, i \int d^{4n+4} x \Box^n J_{\mu}(x) \Box^n T^{\mu}(x) \bigg) \,,
\end{aligned} 
\end{equation}
and the gauge-fixed sourceless generating functional is
\begin{equation} 
\begin{aligned}
Z_{\mathrm{GF}}[0] &= \int \mathcal{D}\Box^n T^{\mu}(x)\,\exp\bigg(\frac{i}{2}\iint d^{4n+4} x \,d^{4n+4} y \,\Box^n T^{\mu}(x)\delta^{4n+4}(x-y)\big(\frac{1}{4^n}\Box \eta_{\mu\nu}\big) \Box^n T^{\nu}(y) \bigg) \\
&=\lim_{N \rightarrow \infty}\sqrt{\frac{4^n(2\pi i)^N}{\mathrm{det}\,[ \delta^{4n+4}(x-y)\Box \eta_{\mu\nu}]}} \,.
\end{aligned}
\end{equation}
The gauge-fixed normalized generating functional is
\begin{equation}
\begin{aligned}
\mathcal{Z}_{\mathrm{GF}}[\Box^n J^{\mu}(x)] &=\frac{Z_{\mathrm{GF}}[\Box^n J^{\mu}(x)]}{Z_{\mathrm{GF}}[0]}\\
&=\exp\bigg(-\frac{i}{2} \iint d^{4n+4}x\,d^{4n+4}y\,\Box^n J^{\mu}(x) (\hat{M}^{-1})_{\mu\nu}(x,y) \Box^n J^{\nu}(y)\bigg) \,.
\end{aligned}
\end{equation}
The free $2$-point correlation function (Feynman propagator) is given by
\begin{equation}
\begin{aligned}
\langle 0 | \mathrm{T} \Box^n \hat{T}_{\nu_1}(x)  \Box^n\hat{T}_{\nu_2}(y)  |0 \rangle &= \frac{1}{i^2} \frac{\delta^n}{\delta\Box^n J^{\nu_1}(x) \delta \Box^n J^{\nu_2}(y) } \mathcal{Z}_{\mathrm{GF}}[\Box^n J^{\mu}] \bigg\vert_{\Box^n J^{\mu} =0} \\
&=i(\hat{M}^{-1})_{\nu_1\nu_2}(x,y) \\
&\equiv B_{\nu_1\nu_2}(x-y) \,.
\end{aligned}
\end{equation}
To find the explicit form of the Feynman propagator, it is more convenient to work in the momentum space. First consider the fourier transform of the vector field $T^{\mu}$
\begin{equation}
T^{\mu}(x) = \int \frac{d^4 p}{(2\pi)^4} \tilde{T}^{\mu}(p) e^{-ip\cdot x} \,.
\end{equation} 
As the $n^{\mathrm{th}}$ order of d'Alembert operator can be expressed as
\begin{equation}
\Box^n = \eta^{\mu_1 \nu_1}\eta^{\mu_2 \nu_2}\cdots \eta^{\mu_n \nu_n} \partial_{\mu_1}\partial_{\nu_1}\partial_{\mu_2}\partial_{\nu_2}\cdots \partial_{\mu_n}\partial_{\nu_n} \,,
\end{equation}
Then we have
\begin{equation}
\begin{aligned}
\Box^n T^{\mu}(x) &=  \int \frac{d^{4n+4} p}{(2\pi)^4}\tilde{T}^{\mu}(p) \eta^{\mu_1 \nu_1}\eta^{\mu_2 \nu_2}\cdots \eta^{\mu_n \nu_n} \partial_{\mu_1}\partial_{\nu_1}\partial_{\mu_2}\partial_{\nu_2}\cdots \partial_{\mu_n}\partial_{\nu_n} e^{-ip_{\alpha} x^\alpha } \\
&=\int \frac{d^{4n+4} p}{(2\pi)^4} (-1)^n p^{2n} \tilde{T}^{\mu}(p) e^{-ip\cdot x}\,.
\end{aligned} 
\end{equation}
Similarly,
\begin{equation}
\Box^n T^{\nu}(x) = \int \frac{d^{4n+4} q}{(2\pi)^4} (-1)^n q^{2n} \tilde{T}^{\mu}(q) e^{-iq\cdot x}\,.
\end{equation}
Then by equation (\ref{eq:step}), the gauge-fixed action in momentum space is
\begin{equation}
\begin{aligned}
S_{\mathrm{GF}} &= \frac{1}{2}\int d^{4n+4} x \Box^n T^{\mu}(x)\big(\frac{1}{4^n}\Box \eta_{\mu\nu} \big) \Box^n T^{\nu}(x) \\
&=  \frac{1}{2}\int d^{4n+4} x \int \frac{d^{4n+4} p \, d^{4n+4} q}{(2\pi)^4 (2\pi)^4} (-1)^{2n} p^{2n}  \tilde{T}^{\mu}(p)\big( -\frac{1}{4^n}q^2 \eta_{\mu\nu}\big) q^{2n}\tilde{T}^{\nu}(q) e^{-i(p+q)\cdot x}\\
&=\frac{1}{2}\int \frac{d^{4n+4}q}{(2\pi)^4} q^{2n}\tilde{T}^{\mu}(-q) \big( -\frac{1}{4^n}q^2 \eta_{\mu\nu}\big) q^{2n}\tilde{T}^{\mu}(q) \\
&=\frac{1}{2}\int \frac{d^{4n+4}p}{(2\pi)^4} p^{2n}\tilde{T}^{\mu}(-p) \tilde{M}_{\mu\nu} p^{2n}\tilde{T}^{\mu}(p) \,,
\end{aligned}
\end{equation}
where
\begin{equation}
\tilde{M}_{\mu\nu} =-\frac{1}{4^n}p^2 \eta_{\mu\nu}\,.
\end{equation}
The gauge-fixed sourced generating functional in momentum space is
\begin{equation}
\begin{aligned}
Z_{\mathrm{GF}} [p^{2n}\tilde{J}_{\mu}(p)] &= \int \mathcal{D}[p^{2n}\tilde{T}^{\mu}(p)] \, \exp\bigg(\frac{i}{2}\int \frac{d^{4n+4} p}{(2\pi)^{4n+4}} p^{2n}\tilde{T}^{\mu}(-p)\,\tilde{M}_{\mu\nu} p^{2n}\tilde{T}(p) \\
&\quad\quad\quad\quad\quad\quad\quad\quad\quad\quad\quad\quad\quad\quad\quad+ i\int \frac{d^{4n+4} p}{(2\pi)^{4n+4}} p^{2n}\tilde{J}_{\mu}(p) p^{2n}\tilde{T}^{\mu}(p)\bigg)
\end{aligned}
\end{equation}
and the sourceless one is
\begin{equation}
Z_{\mathrm{GF}} [0] = \int \mathcal{D}[p^{2n}\tilde{T}^{\mu}(p)] \, \exp\bigg(\frac{i}{2}\int \frac{d^{4n+4} p}{(2\pi)^{4n+4}} p^{2n}\tilde{T}^{\mu}(-p)\,\tilde{M}_{\mu\nu} p^{2n}\tilde{T}(p) \bigg) \,.
\end{equation}
The gauge-fixed normalized generating functional is
\begin{equation}
\mathcal{Z}_{\mathrm{GF}}[p^{2n}\tilde{J}^{\mu}(p)] =\exp\bigg(-\frac{i}{2} \iint \frac{d^{4n+4} p \,d^{4n+4} q}{(2\pi)^{4n+4} \,(2\pi)^{4n+4}} p^{2n}\tilde{J}^{\mu}(p)\,[(M^{-1})_{\mu\nu} (p,q)] q^{2n}\tilde{J}^{\nu}(q) \bigg) \,,
\end{equation}
where the Feynman propagator in momentum space is given by
\begin{equation}
\tilde{B}_{\mu\nu}(p,q) = i(M^{-1})_{\mu\nu} (p,q)\,.
\end{equation}
Since the propagator satisfies
\begin{equation} 
\int \frac{d^{4n+4} k}{(2\pi)^{4n+4}} (\tilde{M}^{-1})^{\mu\rho}(p,k) \, \tilde{M}_{\nu\rho}(k,q)= (2\pi)^4 \delta^{\mu}_{\rho} \delta^{4n+4}(p-q) \,, 
\end{equation}
it simply follows that
\begin{equation}
\begin{aligned}
(\tilde{M}^{-1})^{\mu\nu} \tilde{M}_{\nu\rho} &= \delta^{\mu}_{\rho} \\
(\tilde{M}^{-1})^{\mu\nu}  \,\big(-\frac{1}{4^n}p^2\eta_{\nu\rho} \big) &= \delta^{\mu}_{\rho} \\
 (\tilde{M}^{-1})^{\mu}_{\,\,\,\rho} &= -\frac{ 4^n \delta^{\mu}_{\rho}}{p^2}\\
 \eta_{\beta\mu} (\tilde{M}^{-1})^{\mu}_{\,\,\,\rho} &= -\frac{ 4^n \eta_{\beta\mu}\delta^{\mu}_{\rho}}{p^2}\\
 (\tilde{M}^{-1})_{\beta\rho} &= -\frac{4^n\eta_{\beta\rho}}{p^2} \,. \\
 \end{aligned}
\end{equation}
Therefore, the Feynman propagator in momentum space is
\begin{equation}
\tilde{B}_{\mu\nu} =  -i\frac{4^n\eta_{\mu\nu}}{p^2} 
\end{equation}
We find 
\begin{equation}
\tilde{B}_{\mu\nu} = 4^n \tilde{D}_{\mu\nu} \,,
\end{equation}
where $\tilde{D}_{\mu\nu}= -\frac{i\eta_{\mu\nu}}{p^2}$ is the normal photon propagator. In position space, therefore we have the two point correlation function as
\begin{equation}
\langle 0 | \mathrm{T} \Box^n \hat{T}_{\mu}(x)  \Box^n\hat{T}_{\nu}(y)  |0 \rangle = B_{\mu\nu}(x-y) = \int \frac{d^{4n+4} p}{(2\pi)^{4n+4}} \frac{-i4^n\eta_{\mu\nu}}{p^2} e^{-ip\cdot (x-y)} \,.
\end{equation}
When $n=0$, this restores us the photon propagator in 4D spacetime.

\section{Generalized Proca action under rotor model}
The Proca action describes a massive gauge field theory, where the extra mass term breaks gauge invariance \citep{proca}. The generalized Proca action under model can be formulated as
\begin{equation}
S_{\mathrm{proca}}= \int d^{4n+4} x \bigg(-\frac{1}{4^{n+1}}  \Box^n G_{\mu\nu} \Box^n G^{\mu\nu} + \frac{1}{2\cdot 4^n} M^2 \Box^n T_{\mu} \Box^n T^{\mu} \bigg)\,.
\end{equation} 
Note that when $n=0$, it gives use back the normal Proca action. The action is not invariant under higher order gauge transformation,
\begin{equation}
\Box^n T_{\nu}^\prime = \Box^n T_{\nu} + \partial_{\nu}\Box^n  \theta (x)\,.
\end{equation}

Now using integration by parts, then the generalized Proca action becomes
\begin{equation}
S_{\mathrm{proca}}= \frac{1}{2}\int d^{4n+4} x \Box^n T^{\mu}(x)\bigg(\frac{1}{4^n} \Big((\Box + M^2 )\eta_{\mu\nu}-\partial_{\mu} \partial_{\nu} \Big) \bigg) \Box^n T^{\nu} \,.
\end{equation}
We have the matrix element as
\begin{equation}
\hat{L}_{\mu\nu} =\frac{1}{4^n} \Big((\Box + M^2 )\eta_{\mu\nu}-\partial_{\mu} \partial_{\nu} \Big)\,.
\end{equation}
In momentum space,
\begin{equation} \label{eq:mom}
\tilde{L}_{\mu\nu} =\frac{1}{4^n} \Big( -(p^2 - M^2 ) \eta_{\mu\nu} + p_{\mu} p_{\nu} \Big) \,.
\end{equation}
To find the Feynman propagator we have to compute the inverse $(\hat{L}^{-1})_{\mu\nu}$. This time unlike the projection tensor case, $(\hat{L}^{-1})_{\mu\nu}$ exists and we do not need to fix the gauge. 
Using the same technique as in section 2, we have the normalized partition as
\begin{equation}
\begin{aligned}
\mathcal{Z}_{\mathrm{proca}}[\Box^n J^{\mu}(x)] &=\frac{Z_{\mathrm{proca}}[\Box^n J^{\mu}(x)]}{Z_{\mathrm{proca}}[0]}\\
&=\exp\bigg(-\frac{i}{2} \iint d^{4n+4}x\,d^{4n+4}y\,\Box^n J^{\mu}(x) (\hat{L}^{-1})_{\mu\nu}(x,y) \Box^n J^{\nu}(y)\bigg) \,.
\end{aligned}
\end{equation}
where we have
\begin{equation}
Z_{\mathrm{proca}}[0] =\lim_{N \rightarrow \infty}\sqrt{\frac{4^n(2\pi i)^N}{\mathrm{det}\,[ \delta^{4n+4}(x-y)\big((\Box + M^2) \eta_{\mu\nu} - \partial_{\mu}\partial_{\nu}\big)]}}
\end{equation}
The free $2$-point correlation function (Feynman propagator) is given by
\begin{equation}
\begin{aligned}
\langle 0 | \mathrm{T} \Box^n \hat{T}_{\nu_1}(x)  \Box^n\hat{T}_{\nu_2}(y)  |0 \rangle &= \frac{1}{i^2} \frac{\delta^n}{\delta\Box^n J^{\nu_1}(x) \delta \Box^n J^{\nu_2}(y) } \mathcal{Z}_{\mathrm{proca}}[\Box^n J^{\mu}] \bigg\vert_{\Box^n J^{\mu} =0} \\
&=i(\hat{L}^{-1})_{\nu_1\nu_2}(x,y) \\
&\equiv A_{\nu_1\nu_2}(x-y) \,.
\end{aligned}
\end{equation}

In momentum space, we have
\begin{equation}
\mathcal{Z}_{\mathrm{proca}}[p^{2n}\tilde{J}^{\mu}(p)] =\exp\bigg(-\frac{i}{2} \iint \frac{d^{4n+4} p \,d^{4n+4} q}{(2\pi)^{4n+4} \,(2\pi)^{4n+4}} p^{2n}\tilde{J}^{\mu}(p)\,[(\tilde{L}^{-1})_{\mu\nu} (p,q)] q^{2n}\tilde{J}^{\nu}(q) \bigg) \,,
\end{equation}
In order to find the inverse, we impose the following as in section 2,
\begin{equation}
(\tilde{L}^{-1})^{\mu\nu} \tilde{L}_{\nu\rho} = \delta^{\mu}_{\rho} \,.
\end{equation}
Using the result in equation (\ref{eq:mom}), one simply finds that the inverse is
\begin{equation}
(\tilde{L}^{-1})_{\mu\nu} = -\frac{4^n}{p^2 - M^2} \bigg( \eta_{\mu\nu} - \frac{p_{\mu}p_{\nu}}{M^2}\bigg)
\end{equation}
We have
\begin{equation}
(\tilde{A}^{-1})_{\mu\nu} = 4^n (\tilde{P}^{-1})_{\mu\nu} \,,
\end{equation}
where $(\tilde{P}^{-1})_{\mu\nu}=-\frac{i}{p^2 - M^2} \big( \eta_{\mu\nu} - \frac{p_{\mu}p_{\nu}}{M^2}\big) $ is the normal Proca propagator.  Therefore we have the two point correlation function of massive abelian gauge field as
\begin{equation}
\langle 0 | \mathrm{T} \Box^n \hat{T}_{\mu}(x)  \Box^n\hat{T}_{\nu}(y)  |0 \rangle = A_{\mu\nu}(x-y) = \int \frac{d^{4n+4} p}{(2\pi)^{4n+4}} \frac{-i4^n}{p^2 - M^2} \bigg( \eta_{\mu\nu} - \frac{p_{\mu}p_{\nu}}{M^2}\bigg) e^{-ip\cdot (x-y)} \,.
\end{equation}
When $n=0$, this give us back the original Feynman propagator for the massive gauge boson in 4D spacetime.

%The gauge-fixed Proca action is given by
%\begin{equation}
%S_{\mathrm{proca,GF}}= \int d^{4n+4} x \bigg(-\frac{1}{4^{n+1}}  \Box^n G_{\mu\nu} \Box^n G^{\mu\nu} -\frac{1}{2 \cdot 4^{n}}( \partial_{\mu}\Box^n T^{\mu})^2 + \frac{1}{2\cdot 4^n} M^2 \Box^n T_{\mu} \Box^n T^{\mu} \bigg)\,.
%\end{equation}

\section{Conclusion}
We have performed the Feynman path integral approach to quantize the generalized abelian gauge field theory in $4n+4$ spacetime dimension and find out the corresponding Feynman propagator, which is a constant scaling of $4^n$ of the original photon propagator. We also employed the same technique to quantize the generalized Proca theory under rotor model, and find out the Feynman propagator for the massive gauge boson case, which turns out to be $4^n$ times the original propagator of the massive gauge boson. In both cases, when $n=0$, these restore back to the original respective Feynman propagators in 4D spacetime.

\end{document}